\documentclass{aa} 
\usepackage{graphicx}
\usepackage{txfonts}
\usepackage{hyperref}
\hypersetup{colorlinks=true, linkcolor=blue, urlcolor=blue}
\usepackage{amssymb}
\usepackage{amsmath}
\usepackage{natbib}
\usepackage{multirow}
\usepackage[flushleft]{threeparttable}
\begin{document} 

\title{Generic profile of a long-lived corotating interaction region and associated recurrent Forbush decrease}
\author{M. Dumbovi\'c\inst{1}
\and B. Vr\v{s}nak\inst{1}
\and M. Temmer\inst{2}
\and B. Heber\inst{3}
\and P. K\"uhl\inst{3}
}
\institute{University of Zagreb, Faculty of Geodesy, Hvar Observatory, Zagreb, Croatia\\
              \email{mdumbovic@geof.hr}
              \email{bvrsnak@geof.hr}
\and University of Graz, Institute of Physics, Graz, Austria\\
\and Christian-Albrechts University in Kiel, Department of Extraterrestrial Physics, Kiel, Germany}
%
\abstract 
 {Corotating interaction regions (CIRs), formed by the interaction of slow solar wind and fast streams that originate from coronal holes (CHs), produce recurrent Forbush decreases, which are short-term depressions in the galactic cosmic ray (GCR) flux.}
 {Our aim is to prepare a reliable set of CIR measurements to be used as a textbook for modeling efforts. For that purpose, we observe and analyse a long-lived CIR, originating from a single CH, recurring in 27 consecutive Carrington rotations 2057-2083 in the time period from June 2007 - May 2009. }
 {We studied the in situ measurements of this long-lived CIR as well as the corresponding depression in the cosmic ray (CR) count observed by SOHO/EPHIN throughout different rotations. We performed a statistical analysis, as well as the superposed epoch analysis, using relative values of the key parameters:\ the total magnetic field strength, $B$, the magnetic field fluctuations, $dBrms$, plasma flow speed, $v$, plasma density, $n$, plasma temperature, $T$, and the SOHO/EPHIN F-detector particle count, and CR count.}
 {We find that the mirrored CR count-time profile is correlated with that of the flow speed, ranging from moderate to strong correlation, depending on the rotation. In addition, we find that the CR count dip amplitude is correlated to the peak in the magnetic field and flow speed of the CIR. These results are in agreement with previous statistical studies. Finally, using the superposed epoch analysis, we obtain a generic CIR example, which reflects the in situ properties of a typical CIR well.}
 {Our results are better explained based on the combined convection-diffusion approach of the CIR-related GCR modulation. Furthermore, qualitatively, our results do not differ from those based on different CHs samples. This indicates that the change of the physical properties of the recurring CIR from one rotation to another is not qualitatively different from the change of the physical properties of CIRs originating from different CHs. Finally, the obtained generic CIR example, analyzed on the basis of  superposed epoch analysis, can be used as a reference for testing future models.}
\maketitle
%
\section{Introduction}
\label{intro}

The interaction between the fast component of the solar wind (high speed streams, HSS) originating from coronal holes (CHs) and the ambient slower solar wind, leads to the creation of the so-called stream interaction regions (SIRs). As CHs are rather long-lived structures, these interaction regions may persist for several solar rotations corotating with the Sun and, therefore, they are called corotating interaction regions (CIRs). The interaction forms a region of compressed plasma around the leading edge of the stream interface (i.e., increased total magnetic field and density), which is followed by a region of high temperature and flow speed \citep[e.g.,][and references therein]{gosling99,jian06b,richardson18}. In their interaction with the galactic cosmic rays (GCRs), CIRs produce so-called recurrent Forbush decreases (FDs), which are short-term depressions in the cosmic ray (CR) count  recurring over several solar rotations,  identically
to CIRs \citep[see e.g., overview by][]{richardson04}. This effect is similar to FDs produced by interplanetary coronal mass ejections \citep[see overview by][]{cane00}. However, comparative studies have shown that the GCR-effectiveness caused by CIRs tends to be much lower than that of an ICME. As measured by neutron monitors, CIR-related FDs rarely have magnitudes larger than 3\%, whereas ICME-related FDs can easily have magnitudes of 10\% and more \citep[e.g.,][]{dumbovic12b,badruddin16,melkumyan19}.

The interaction of GCRs with interplanetary plasma and magnetic field structures can be described by the transport theory, with GCRs experiencing diffusion, drifts, convection, and energy change \citep[e.g.,][]{parker65}. Statistical studies have shown that there is a strong anti-correlation between the solar wind speed and the CR count time-series during a CIR \citep[e.g.,][]{richardson96}, favoring convection as the main transport mechanism. On the other hand, the recurrent FD magnitude was found to be correlated to the magnitude of the total magnetic field in the CIR  \citep[e.g.,][]{calogovic09}, indicating diffusion as an important transport mechanism. Such correlation was not found by \citet{richardson96}, for instance, or in a later study by \citet{dumbovic12b}. However, a more recent statistical study by \citet{melkumyan19} revealed that the recurrent FD magnitude was moderately correlated to both plasma flow speed magnitude and the magnitude of the total magnetic field in the CIR, favoring both convection and diffusion as transport mechanisms. These inconsistencies between different studies (or different samples used in different studies) may hamper modeling efforts and lead to inconclusive results. In their majority, studies seem to favor the combined convection-diffusion approach for explaining the CIR-related GCR modulation. However, these studies were based on samples containing CIRs formed by HSS originating from different CHs, throughout different time-frames.

It is reasonable to assume that the physical properties of different CHs may be transferred differently into the heliosphere. Indeed, it was found that for the well established CH area -- HSS plasma flow speed peak relation  \citep[e.g.,][]{nolte76,vrsnak07a,tokumaru17,hofmeister18},  the correlation coefficients and slopes of the linear regression line are varying between different long-lived CHs samples \citep{heinemann20}. Therefore, we might expect that a single, long-lived CIR originating from the same CH might represent a ``cleaner" sample compared to CIRs corresponding to different CHs. In the current study, we focus on a single CIR example aiming to be used as a reference for modeling efforts \citep[companion paper by][]{vrsnak22}. For that purpose, we observe and analyse a long-lived CIR, associated with a single CH recurring in 27 consecutive Carrington rotations in the time-period 2007--2009, around the minimum activity of solar cycle 24. This time period was characterized by very low CME activity and thus provided good conditions for studying CHs and their related CIRs \citep[e.g.,][]{gomez-herrero09,dresing09,gomez-herrero11,kuhl13}. \citet{gil18} further studied the greatly enhanced recurrence of cosmic rays in this time period. However, none of these studies carried out an individual study of  CIRs originating from specific CHs and their influence on the GCRs.

\section{Data and method}
\label{data}

\begin{figure}
\centering
\includegraphics[width=0.48\textwidth]{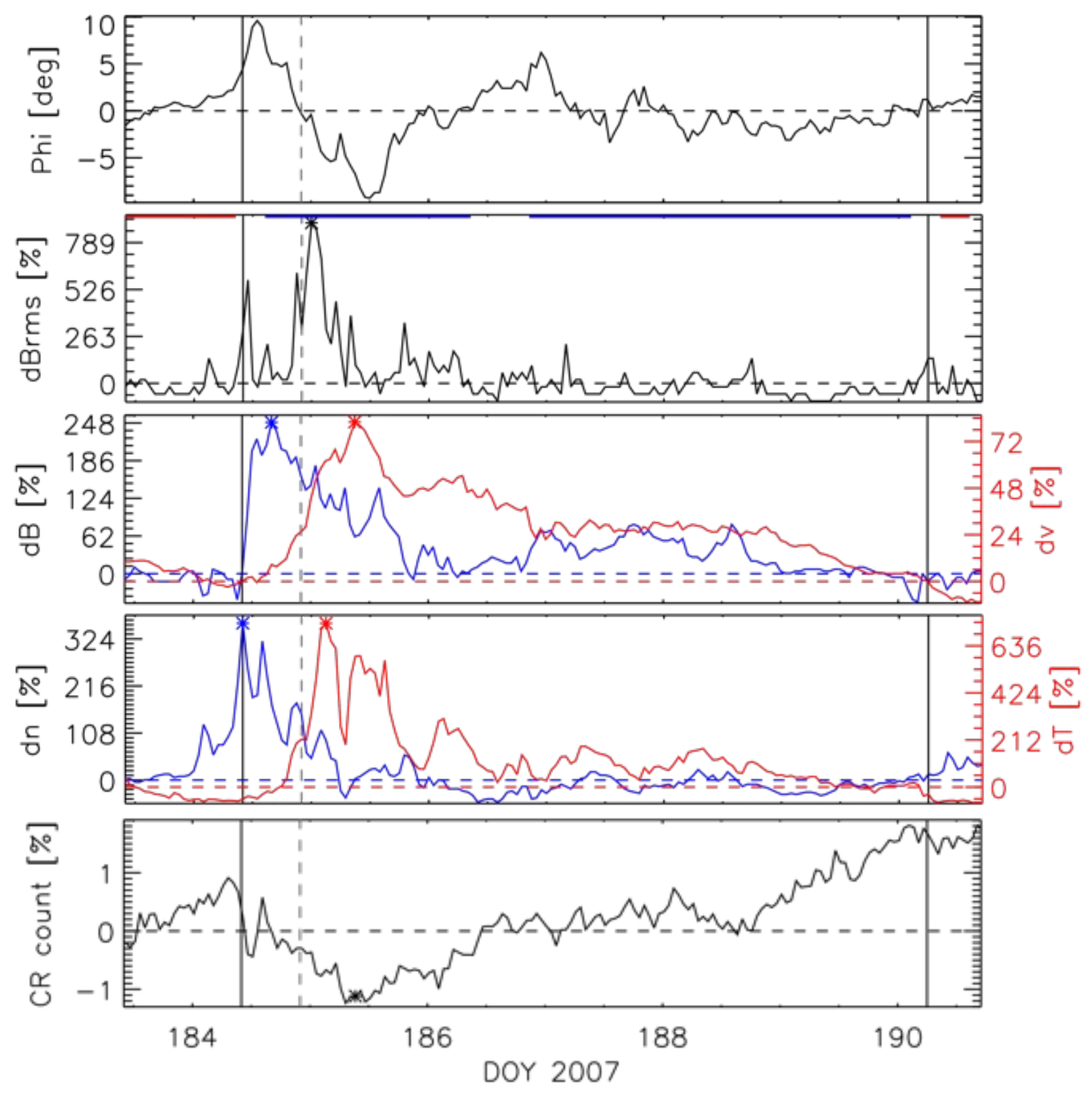}
\caption{In situ measurements for the CIR recorded in Carrington rotation 2058 (rot2). Panels top to bottom show: 1) the azimuth flow angle calculated in the RTN system, $Phi$, 2) the magnetic field fluctuations, $dBrms$, and the in situ magnetic polarity calculated according to \citet{neugebauer02} (red and blue overlying lines for positive and negative polarity, respectively), 3) the total magnetic field strength, $B$, and plasma flow speed, $v$, 4) plasma density, $n$, and plasma temperature, $T$, and 5) the SOHO/EPHIN F-detector particle count, $CR\,\,count$. Vertical solid lines mark the onset and trail edge times, whereas the vertical dashed lines mark the stream interface time. All parameters, except $Phi,$ are given in relative values. Horizontal dashed lines mark referent levels, which were obtained by normalizing to the following values: $Brms_{\mathrm{norm}}=0.25$\,nT, $B_{\mathrm{norm}}=3.0$\,nT, $v_{\mathrm{norm}}=346\,\mathrm{km~s}^{-1}$, 
$n_{\mathrm{norm}}=3.9\,\mathrm{cm}^{-3}$, $T_{\mathrm{norm}}=0.43\cdot10^{5}$\,K, and $CR\,\,count_{\mathrm{norm}}=22940$ (for details see main text). Asterisks mark the measured peak values: $Brms_{\mathrm{peak}}=2.5$\,nT (900\% increase), $B_{\mathrm{peak}}=10.4$\,nT (249\% increase), $v_{\mathrm{peak}}=630\,\mathrm{km~s}^{-1}$ (82\% increase), $n_{\mathrm{peak}}=17.8\,\mathrm{cm}^{-3}$ (360\% increase), $T_{\mathrm{peak}}=3.64\cdot10^{5}$\,K (738\% increase), and $CR\,\,count_{\mathrm{dip}}=22684$ (-1.1\% decrease).}
\label{fig1}
\end{figure}

We investigate the recurring CIR near 1au using 1 hour plasma and magnetic field data in the radial-tangential-normal (RTN) system provided by the OMNI\_M database of the \href{https://cohoweb.gsfc.nasa.gov/coho/html/cw_data.html}{COHOWeb} Service, supplemented by the 1 hour magnetic field fluctuation data (rms; standard deviation in average magnitude) provided by the \href{https://omniweb.gsfc.nasa.gov/html/ow_data.html#norm_pla}{OMNIWeb} database \citep{king05}. The CIR is identified manually by observing the stream interface properties: drop in proton density, rise in proton temperature, change in the azimuthal flow angle and a steepening flow speed-time profile, associated with increased magnetic field and magnetic field fluctuations \citep[e.g.,][and references therein]{gosling99, jian06b, richardson18}. In addition, we use the single detector count rate of detector F of the \textit{Electron Proton Helium Instrument} \citep[EPHIN,][]{muller-mellin95} onboard the \textit{Solar and Heliospheric Observatory} (SOHO). Namely, we use the hourly count rate of the F-detector which is suitable to observe CR flux of energies $>$ 50 MeV \citep{kuhl15}, where a clear recurring depression is observed corresponding to the identified CIR. The OMNI in situ measurements are time-shifted to the Earth's bow shock nose, therefore, we applied the same time shift to EPHIN data using the SOHO spacecraft daily coordinates. This way we make certain that any time shift of the OMNI data compared to EPHIN data is below the chosen hourly data resolution.

\begin{table*}
\caption{Polarity and characteristic timings of CH-CIR throughout different rotations.}
\label{tab1}
\centering
\begin{tabular}{cccccccccccc} 
\hline           
rotation        &       Carrington      &                       &       HSS                     &       HSS             &       CH      &       CH              &       onset   &       SI              &       trailing         &       end\\
number  &       rotation        &       year            &       peak time               &       polarity        &       time    &       polarity        &       time            &       time            &       edge time    &       time\\
\hline
rot1            &       2057            &       2007            &       160.8           &               --      &       157     &       NaN             &       158,8   &       160,4   &       162,7   &       163,2\\
rot2            &       2058            &       2007            &       185.4           &               --      &       181     &       NaN             &       184,4   &       184,9   &       190,3   &       190,7\\
rot3            &       2059            &       2007            &       213.8           &               --      &       207     &       --              &       210,0   &       210,3   &       216,8   &       217,3\\
rot4            &       2060            &       2007            &       239.8           &               --      &       235     &       --              &       237,4   &       237,8   &       NaN             &       243,9\\
rot5            &       2061            &       2007            &       267.3           &               --      &       262     &       --              &       263,4   &       263,9   &       NaN             &       270,5\\
rot6            &       2062            &       2007            &       293.0           &               --      &       289     &       --              &       291,0   &       291,4   &       NaN             &       295,5\\
rot7            &       2063            &       2007            &       318.4           &               --      &       316     &       --              &       316,7   &       317,2   &       322,5   &       323,3\\
rot8            &       2064            &       2007            &       345.3           &               --      &       343     &       --              &       343,0   &       344,8   &       NaN             &       351,0\\
rot9            &       2065            &       2008            &       7.5                     &               --      &       4       &       --              &       4,9             &       5,3             &       NaN             &       12,2    \\
rot10           &       2066            &       2008            &       34.5                    &               --      &       30      &       --              &       31,5            &       31,7            &       NaN             &       38,5    \\
rot11           &       2067            &       2008            &       61.3                    &               --      &       58      &       --              &       58,6            &       59,7            &       NaN             &       65,0    \\
rot12           &       2068            &       2008            &       88.4                    &               --      &       86      &       --              &       86,0            &       86,5            &       93,7            &       94,0    \\
rot13           &       2069            &       2008            &       114.8           &               --      &       112     &       --              &       113,9   &       114,3   &       NaN             &       121,6\\
rot14           &       2070            &       2008            &       142.6           &               --      &       139     &       --              &       140,0   &       140,6   &       NaN             &       149,0\\
rot15           &       2071            &       2008            &       169.1           &               --      &       165     &       --              &       166,6   &       166,7   &       NaN             &       171,8\\
rot16           &       2072            &       2008            &       196.3           &               --      &       193     &       --              &       193,1   &       193,6   &       NaN             &       202,5\\
rot17           &       2073            &       2008            &       222.8           &               --      &       220     &       --              &       222,0   &       222,3   &       229,5   &       229,8\\
rot18           &       2074            &       2008            &       251.3           &               --      &       246     &       --              &       247,0   &       247,4   &       NaN             &       256,8\\
rot19           &       2075            &       2008            &       276.9           &               --      &       273     &       --              &       274,6   &       274,8   &       282,7   &       284,3\\
rot20           &       2076            &       2008            &       303.7           &               --      &       301     &       --              &       302,2   &       302,9   &       309,5   &       311,0\\
rot21           &       2077            &       2008            &       331.9           &               --      &       327     &       --              &       330,0   &       330,2   &       336,4   &       337,5\\
rot22           &       2078            &       2008            &       358.5           &               --      &       355     &       --              &       357,0   &       358,0   &       365,0   &       365,5\\
rot23           &       2079            &       2009            &       19.9                    &               --      &       15      &       --              &       18,2            &       19,0            &       NaN             &       22,5    \\
rot24           &       2080            &       2009            &       46.4                    &               --      &       42      &       NaN             &       44,8            &       45,3            &       NaN             &       50,2    \\
rot25           &       2081            &       2009            &       73.1                    &               --      &       69      &       --              &       70,9            &       72,2            &       78,1            &       79,0    \\
rot26           &       2082            &       2009            &       101.8           &               --      &       96      &       --              &       98,8            &       99,1            &       NaN             &       105,0\\
rot27           &       2083            &       2009            &       128.2           &               --      &       124     &       NaN             &       126,0   &       126,6   &       132,5   &       133,0\\
\hline
\end{tabular}
\begin{tablenotes}
        \item \small Overview of all rotations of the recurring CIR and its associated CH, its polarity, and characteristic timings. Times are given in day of year (DOY). CH time marks the DOY of the passage over the central meridian.
\end{tablenotes}
\end{table*}

The CIR signatures recur in 27 consecutive Carrington rotations 2057--2083 in the time period from June 2007--May 2009 (rot1--rot27 in Table\ref{tab1}). We note that for Carrington rotations preceding rot1 and following rot27, we also observed CIR-like in situ signatures, however, they were weak or inconclusive and thus, we did not consider them in the analysis. It is particularly interesting to note that in the first rotation after the last one studied, peaks of $n$ and $B$, as well as a dip in the CR count, are still fully recognizable although there was no signature of a HSS, that is, no increase of the flow velocity is present. This implies that the compression and the resulting CIR signature are still slowly decaying even after the HSS has already decayed. In order to confirm that the observed recurring CIR, namely, the corresponding HSS, originates from the same recurring CH, we make the HSS-CH association using the timing and polarity criteria, similarly to what was done in \citet{heinemann20}. The timing criterion follows the statistical results presented by \citet{vrsnak07a}, where the average delay of the HSS peak to the time of a CH passing the central meridian was found to be $3.6\pm0.7$ days. The polarity criterion imposes that the polarity of the CH should correspond to the polarity of the HSS. We calculated the polarity of the HSS following Eq. 1 from \citet{neugebauer02}, where the magnetic polarity, $P$, is the cosine of the angle between the average field direction in the RT plane and the expected direction of an outward directed Parker spiral.

The polarity of CHs was obtained from full Sun drawings using UV, X-ray, or He I 10830 \AA\, chromospheric line observations, available at the National Oceanic and Atmospheric Administration (NOAA) Solar Data Services under \href{https://www.ngdc.noaa.gov/stp/solar/solar-imagery.html}{``solar imagery,"} for manually derived coronal holes. The corresponding information are presented in Table \ref{tab1}. For 4 rotations there were no appropriate reports from the NOAA database. Therefore, we supplement the NOAA catalogue based on CH observations from the \textit{Extreme-ultraviolet Imaging Telescope} \citep[EIT,][]{delaboudiniere95} onboard SOHO using the \href{https://www.jhelioviewer.org}{JHelioviewer} service \citep{muller17}, and from the \textit{X-ray Telescope} \citep[XRT,][]{golub07} onboard Hinode spacecraft \citep{kosugi07} using the \href{https://www.solarmonitor.org}{Solar Monitor} service. With EIT and XRT observations, we confirm the day of the CH passage over the central meridian for four rotations missing in the NOAA database. In addition, for these four rotations we use the Collection of Analysis Tools for Coronal Holes (CATCH) tool \citep{heinemann19b} to determine the polarity of the CH. For rot1 and rot2, the results indicate weakly unipolar magnetic field of negative polarity (with high uncertainty), whereas for the last two rotations, the results are inconclusive. Therefore, since we do not have reliable polarity information for
these four rotations, we rely on the timing criterion.

For the time intervals of interest, we calculate the relative values of the following parameters: the azimuth flow angle calculated in the RTN system: $Phi$; the magnetic field fluctuations: $dBrms$; the total magnetic field strength: $B$; plasma flow speed: $v$; plasma density: $n$; plasma temperature: $T$; and the SOHO/EPHIN F-detector particle count: $CR\,\,count$. Figure~\ref{fig1} summarizes these parameters for Carrington rotation 2058 (rot2). The beginning of the analyzed time-interval is defined based on the onset time of each CIR event. The onset time is determined manually by the observer as the start of the plasma flow speed increase which roughly marks the ``reach'' of the HSS influence at the Lagrangian point 1 (L1). In theory, we expect the HSS to accelerate the plasma ahead, thus we expect to observe the increasing speed profile in the slow wind ahead of the HSS. The point where we observe such behavior is designated as the onset time. The beginning of each of the under-studied time intervals is defined one day before the onset time of the event. The end of the analyzed time interval is also determined manually by the observer as the point where quiet conditions (pre-event levels) are reached or, as the last point in the descending phase of an identified CIR until the start of another disturbed condition in the solar wind (either another SIR or an ICME). We note that there is some subjectivity in determining the onset and end times manually, especially given the small but notable variability in measurements of flow speed. More specifically, the time uncertainty for both the onset and end times of each event, is estimated in the range of hours. However, the manual detection provides a simple and consistent way to determine the time-period of interest, with a start time defined as: start time = onset time $-$ 1 day.

Along with the start, onset, and end times there are two additional timings we determine during the arrival of a CIR at L1: the stream interface and the trailing edge times. These are determined based on strict criteria once the observed time-interval is defined. The stream interface time is determined as the point where $Phi=0$, with $B$, $n$, $v,$ and $T$ showing increasing profiles. The underlying assumption lies in the fact that the stream interface is the point where the HSS meets the slow stream and, as a result, the two streams are deflected. We note that we do not follow the sign convention introduced by \citet{gosling78}, namely, that negative flow angles correspond to flow in the direction of planetary motion about the Sun (westward). On the contrary, we keep the original sign of the azimuth flow angle as calculated in the RTN system (provided by OMNI\_M database). We further determine the trailing edge time as the time where the flow speed in the HSS returns to the value corresponding to the onset time. We note that for a subset of analyzed time intervals, the flow speed never returned to the initial, settled solar wind levels and thus the trailing edge time could not be measured.

To observe changes in the CR counts such as recurrent Forbush decreases, typically relative values are analyzed. We adopt the same methodology for solar wind and interplanetary magnetic field parameters. The pre-event levels of different in situ parameters change from one event to another. The relative change of the in situ parameters may thus be different for the same peak values. Normalizing each parameter to its pre-event level, we obtain relative values, which facilitates comparison of different events. In order to obtain relative values, we normalize the flow speed recorded during each of the under-studied time periods to the value detected at the onset time of each such period. For other parameters (except $Phi$ which is left in absolute values), relative values are obtained by normalizing to the average value in the first 6 hours of the analyzed time period. In some events, trends can be observed in CR count prior to the onset time and thus we expect that different normalization windows (e.g., 3-hours, 6-hours, 12-hours) might yield slightly different relative peaks. By consistently using the same normalization time-window of 6 hours, we minimize this effect. In addition to the length of the normalizing period, selection of the start of the normalizing period can also influence the relative peaks, especially if it is (partly) related to disturbed conditions due to the prior event. Therefore, we choose one day prior to the onset time as the ``optimal distance", since the timing is close enough to the event to avoid using normalization levels of non-quiet time conditions due to prior events (e.g., ICMEs or HSS), but presumably far enough not to catch the start of the event (as seen in different parameters).

Finally, we measure the peak values of the key parameters (except $Phi$). In Fig.~ \ref{fig1}, we see that the peaks of all key parameters are pronounced and well defined, except the CR count. The dip of the latter can be characterized as a plateau with small variations around it. Therefore, the value of the CR count dip amplitude was determined as the average of the minimum of the CR count and its two neighboring values. The normalization values, as well as the relative peak values for all measured key parameters for all rotations are given in Table \ref{tab_app1} in Appendix \ref{app1}.

\section{Results and discussion}
\label{results}
\subsection{CIR evolution and the corresponding CR response}
\label{evolution}


\begin{figure*}
\centering
\includegraphics[width=0.9\textwidth]{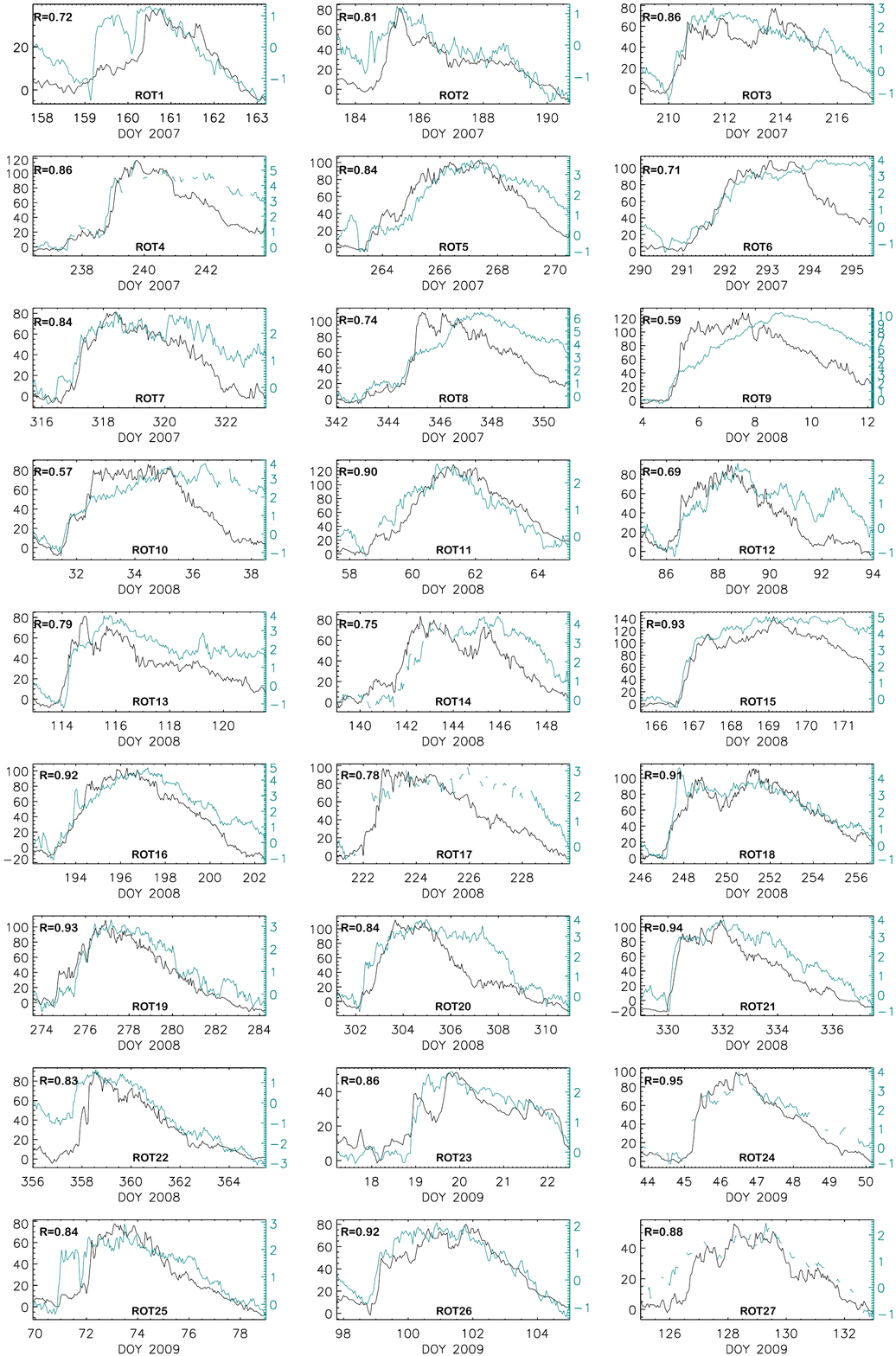}
\caption{Relative values of the flow speed (black) and mirrored $CR\,\,count$ profiles (teal) for 27 rotations of the recurring CIR (rot1--rot27 from top-left to right bottom, respectively). The y-axis units are \% in all subplots for both flow speed and CR counts. The calculated Pearsons correlation coefficient between the two curves, R, are given in the upper left corner of the corresponding subplot. Deviations between the two time-profiles in the rising phase for rotations 1, 18, and 25 are due to ICMEs.}
\label{fig2}
\end{figure*}

We first analyze the evolutionary properties of the CIR plasma flow speed and the corresponding CR count. In Figure \ref{fig2}, we show the time-evolution of the in situ plasma speed of the CIR and the corresponding mirrored CR count throughout the 27 rotations. For each rotation, the calculated Pearsons correlation coefficient between the two is displayed. A striking resemblance of the time-profiles of both parameters' relative values can be seen in each rotation, reflected by correlation coefficients ranging from 0.57 to 0.95 (moderate to strong correlation). This is in agreement with previous studies \citep[e.g.,][]{richardson96}. Notable deviations in the rising phase of the time profiles are observed for rotations 1, 18, and 25 due to the arrival of ICME structures at L1 (the in situ measurements for these rotations are analyzed in more detail and discussed in Appendix \ref{app2}). Another notable difference can be observed in the declining phase of the time-profiles in many rotations (mostly at the end of 2007 and throughout 2008). The CR count time-profile does not recover at the same rate as the speed-time profile. This might be related to another long-living CIR which in many rotations trails our analyzed CIR. However, the correlation between the flow speed and the mirrored CR count-time profiles throughout different rotations is clear regardless of the few deviations and should be captured by any model describing the CIR-GCR interaction.

\subsection{Statistical relation between key parameters and CR counts}
\label{statistics}

\begin{figure}
\centering
\includegraphics[width=0.48\textwidth]{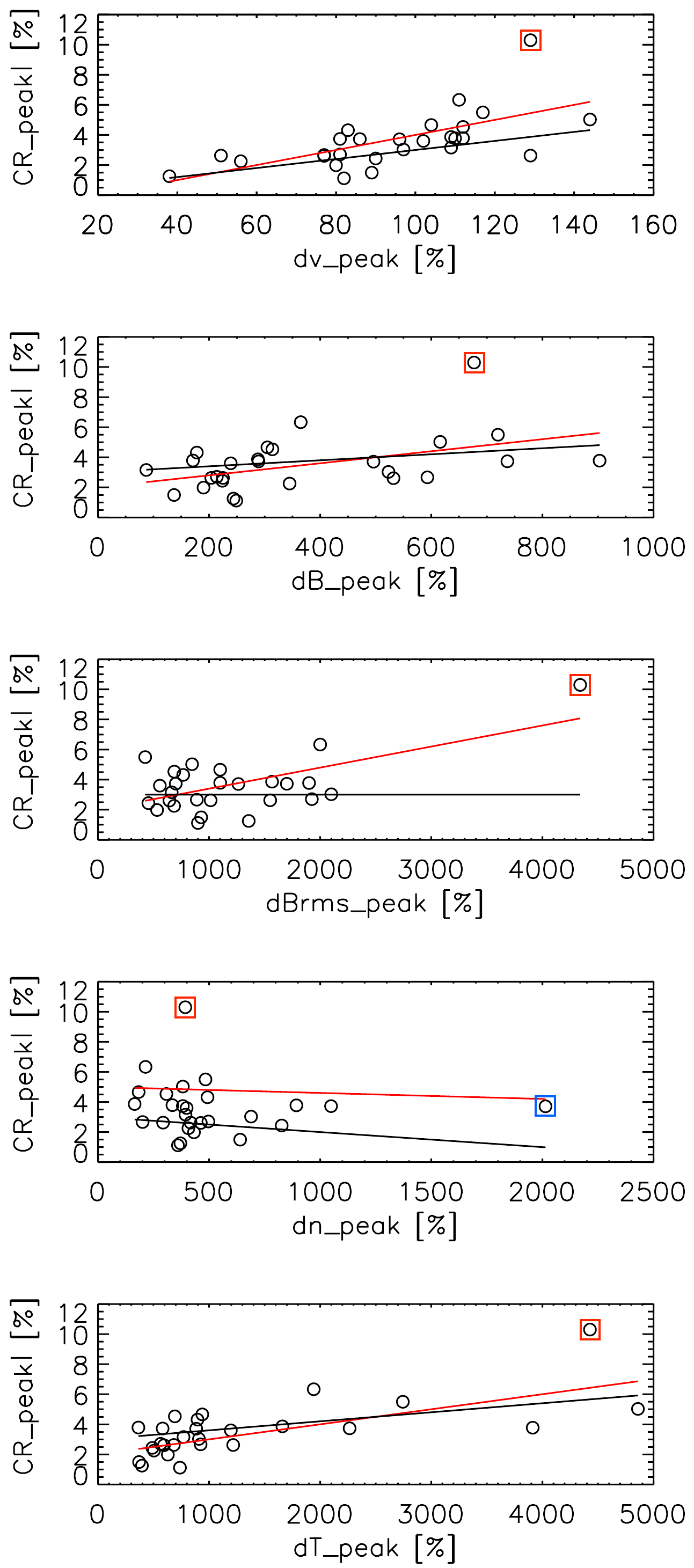}
\caption{Scatter plots of CR count dip amplitudes versus key parameter peak values and corresponding linear fits with outlier (red line) and without outlier (black line) obtained by bootstrapping. The outlier is marked by a red square and the second outlier noticed in density versus CR count scatter plot is marked by a blue square.}
\label{fig3}
\end{figure}

We also performed a statistical analysis between the CR count peak values and the peak values of the key parameters. We analyzed the scatter-plots of all the relative peak values of the key parameters and the mirrored CR count by performing a linear regression fitting to obtain the slope and intercept. We further calculate the Pearson correlation coefficient, $R$ (Fig.\,\ref{fig3}). The same analysis was repeated by using a bootstrap method, which we used to iteratively resample the dataset using random sampling with replacement on the original dataset. This was performed 10,000 times to calculate the median and 95\% confidence interval for the slope, intercept and $R$ (Table \ref{tab2}). A data point can be seen in Fig.\,\ref{fig3} differing significantly from other data points, especially in the $dBrms_\mathrm{peak}$ versus $CR_\mathrm{peak}$ subplot. This outlier is marked by a red square in the same figure and it corresponds to a CIR recorded in rotation 9, which shows not only a large CR count amplitude compared to other rotations, but also an extreme peak in $Brms$ (shown and discussed in the Appendix \ref{app3}). Since $R$ is sensitive to outliers, we repeat the analysis for a sample where the outlier (outlined red in Fig.\,\ref{fig3}) is excluded. Moreover, in the scatter-plot showing the density and CR count relative peak values we find an additional outlier (CIR recorded in rotation 24, marked in blue in Fig.\,\ref{fig3}). However, this outlier does not stand out in any other scatterplots, therefore, we do not exclude it from statistics of parameters other than plasma density. As shown and discussed in the Appendix \ref{app3}, other than the unusually large density peak, the event is quite ordinary.

We can see that there is a moderate correlation between CR count dip amplitudes and relative peak values of plasma flow speed and temperature (R=0.6, see Table \ref{tab2}), which does not change when the outlier is removed (R=0.6, see Table \ref{tab2}). There is also a moderate correlation between CR count dip amplitudes and the relative peak of the magnetic field (R=0.4, see Table \ref{tab2}), which does not change with the removal of the outlier (R=0.4, see Table \ref{tab2}). On the other hand, the correlation between the CR count dip amplitude and the $Brms$ (R=0.6, see Table \ref{tab2}) is caused by the outlier. The correlation is completely lost when this outlier is removed (R=0.1, see Table \ref{tab2}). We note that in this relatively small and ``clean" sample (single recurrent CH-CIR), we can easily identify and analyze the outlier to obtain reliable correlation analysis. In larger samples, containing a variety of different CH-CIR pairs, identifying such outliers might not be as simple. The latter might explain some inconsistencies between different statistical studies.

The statistical analysis performed here shows that CR count dip amplitudes are correlated with relative peak values of plasma flow speed, in agreement with previous studies by \citet{richardson96} and \citet{melkumyan19}. In addition, the analysis shows that CR count dip amplitudes are correlated with relative peak values of the total magnetic field, in agreement with previous studies by \citet{calogovic09} and \citet{melkumyan19}. Our results therefore indicate that both convection and diffusion are relevant mechanisms for producing recurrent FDs. In our study we use a ``clean'' sample based on a long-lived CIR, associated with a single CH. However, we note that qualitatively, our statistical results do not differ from those using samples related to different CHs \citep{richardson96,calogovic09,melkumyan19}. This indicates that the change of the physical properties of the recurring CIR from one rotation to another is qualitatively no different from the change of the physical properties of CIRs originating from different CHs.

\begin{table}
\caption{Bootstrapping results}
\fontsize{8}{20}\selectfont
\label{tab2}
\centering
\begin{threeparttable}
\begin{tabular}{|c|c|c|c|c|} 
\hline           
                                                                                        & parameter       & slope                                         & intercept                 & $R$\\
\hline
\multirow{5}{*}{\rotatebox[origin=c]{90}{with outlier}}         & $v$           & $0.05^{+0.04}_{-0.02}$                  & $-1^{+2}_{-3}$        & $0.6^{+0.1}_{-0.2}$\\
\cline{2-5}
                                                                                        & $B$             & $0.004^{+0.05}_{-0.03}$               & $2^{+1}_{-1}$         & $0.4^{+0.2}_{-0.3}$\\
\cline{2-5}
                                                                                        & $Brms$          & $0.0014^{+0.0006}_{-0.0019}$  & $2^{+2}_{-1}$         & $0.6^{+0.3}_{-0.8}$\\
\cline{2-5}
                                                                                        & $n$             & $-0.0004^{+0.0009}_{-0.0029}$ & $4^{+5}_{-3}$         & $-0.1^{+0.2}_{-0.2}$\\
\cline{2-5}
                                                                                        & $T$             & $0.0010^{+0.0009}_{-0.0006}   $       & $2^{+1}_{-1}$         & $0.7^{+0.2}_{-0.3}$\\
\hline
\multirow{5}{*}{\rotatebox[origin=c]{90}{without outlier}}      & $v$           & $0.03^{+0.02}_{-0.02}$                  & $0^{+1}_{-2}$         & $0.6^{+0.2}_{-0.3}$\\
\cline{2-5}
                                                                                        & $B$             & $0.002^{+0.002}_{-0.002}$             & $3^{+3}_{-2}$         & $0.4^{+0.2}_{-0.3}$\\
\cline{2-5}
                                                                                        & $Brms$          & $0.000^{+0.001}_{-0.001}$             & $3^{+1}_{-1}$         & $0.1^{+0.4}_{-0.4}$\\
\cline{2-5}
                                                                                        & $n$\tnote{*}    & $-0.001^{+0.001}_{-0.002}$            & $3^{+1}_{-1}$         & $-0.1^{+0.3}_{-0.3}$\\
\cline{2-5}
                                                                                        & $T$             & $0.0006^{+0.0010}_{-0.0003}   $       & $3^{+1}_{-1}$         & $0.6^{+0.2}_{-0.2}$\\
\hline
\end{tabular}
\begin{tablenotes}
\tiny
        \item The slope, intercept and the Pearson correlation coefficient, $R$, for scatter-plots of relative peak values and mirrored CR count peak in Figure \ref{fig3} obtained by bootstrapping with and without the outlier.
        \item[*] \textit{2 outliers were identified and removed}
\end{tablenotes}
\end{threeparttable}
\end{table}

\begin{figure*}
\centering
\includegraphics[width=0.8\textwidth]{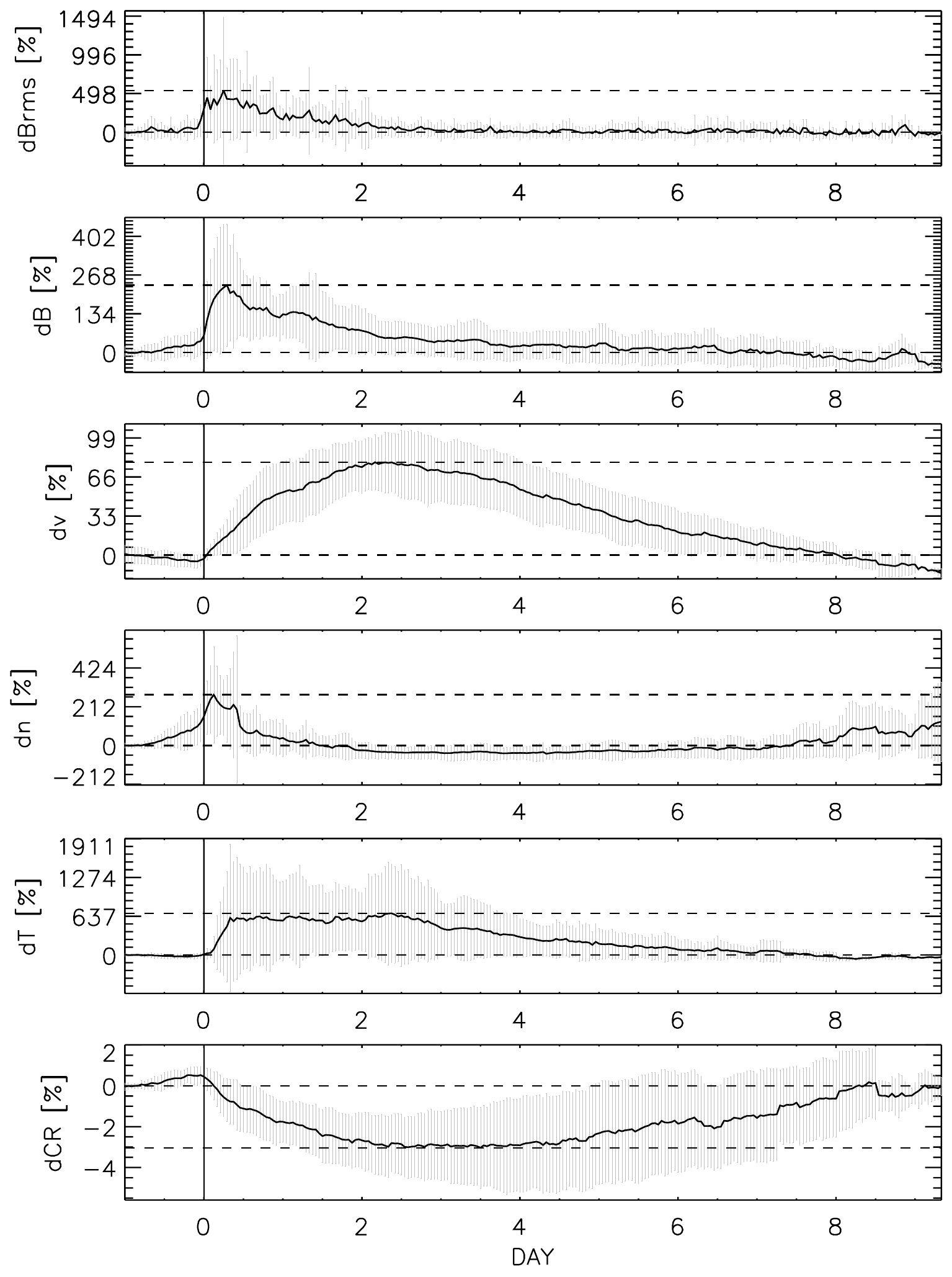}
\caption{SEA curves (black) with standard deviations (gray). Panels, top to bottom, show: 1) magnetic field fluctuations, $dBrms$, 2) total magnetic field strength, $B$, 3) plasma flow speed, $v$, 4) plasma density, $n$, 5) plasma temperature, $T$, and 6) SOHO/EPHIN F-detector particle counts, $CR\,\,count$. Vertical solid line marks the zero epoch (the onset time) and two gray, dashed horizontal lines mark 0 and peak values (in \%). All parameters are given in relative values, with the average normalization values (see Appendix \ref{app3}): $Brms_{\mathrm{norm}}=0.22$\,nT, $B_{\mathrm{norm}}=3.1$\,nT, $v_{\mathrm{norm}}=329\,\mathrm{km~s}^{-1}$, $n_{\mathrm{norm}}=4.9,\mathrm{cm}^{-3}$, $T_{\mathrm{norm}}=0.36\cdot10^{5}$\,K, and $CR\,\,count_{\mathrm{norm}}=24412$. The peak values are: $Brms_{\mathrm{peak}}=1.6$\,nT (538\%), $B_{\mathrm{peak}}=10.3$\,nT (233\%), $v_{\mathrm{peak}}=586\,\mathrm{km~s}^{-1}$ (78\%), $n_{\mathrm{peak}}=18.5\,\mathrm{cm}^{-3}$ (278\%), $T_{\mathrm{peak}}=2.82\cdot10^{5}$\,K (684\%), and $CR\,\,count_{\mathrm{dip}}=23680$ (-3.0\%).}
\label{fig4}
\end{figure*}

\subsection{Superposed epoch analysis}
\label{sea}

\begin{figure*}
\centering
\includegraphics[width=0.8\textwidth]{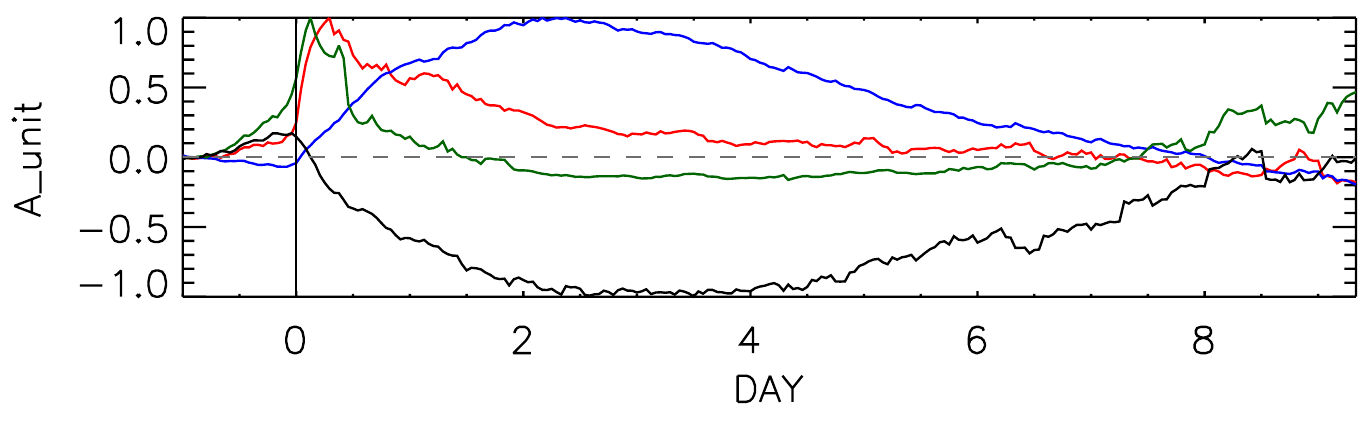}
\caption{SEA curves for different parameters: CR count (black), flow speed (blue), magnetic field strength (red), and density (green). Vertical solid line marks the zero epoch (the onset time) and gray, dashed horizontal line marks 0. The peak values of all SEA curves have been normalized to 1.}
\label{fig5}
\end{figure*}

In order to characterize a text-book CIR for modeling purposes, we performed the superposed epoch analysis (SEA) using relative values of the key parameters. The onset time was taken as the zero-epoch (i.e., the point of reference for superposition). The main benefit of the SEA is that it visualizes the general shape and the relative timings of the change in different parameters in a CIR. We do not consider stream interface timing as the zero epoch, since the time period from the onset time to stream interface may differ substantially from one event to another (from 0.1 day to 1.8 days, see Table\,\ref{tab1}).

ICME signatures might substantially disturb the general shape of the CIR and timings of different sub-regions within the CIR (compression region, stream interface, HSS), therefore, they were excluded from the dataset (rotations 1, 18, and 25). We do not remove rot9 and rot24 from the SEA sample. As discussed in Appendix \ref{app3}, it is possible that they only appear as outliers in their respective scatter plots due to normalization or the extreme peak value, both of which are annulled by the averaging process in the frame of the SEA. Finally, some concern arises regarding the CR count data gaps observed in some of the events (see Fig.\,\ref{fig2}) and how they might influence SEA. We find that in all except four events (rot4, rot17, rot24, and rot27) less than 25\% of data is missing in the observed time interval. Moreover, even when data gaps exist, the main trend of the curve is still well visible, as can be seen in Fig.\,\ref{fig2}. The data gaps are not very long, therefore, the average curves should still represent the events rather well. We repeat SEA on a sample where these four events are excluded (rot4, rot17, rot24, and rot27) and obtain almost identical results to SEA on a sample where the four events are not excluded. Therefore, we performed a SEA analysis using a sample which includes these four events. In Fig.\,\ref{fig4} we can see that the CIR related recurrent FD shows a shallow and symmetric profile, as expected based on previous studies using SEA \citep[e.g.][]{badruddin96, dumbovic12b, kumar14}. 

As can be seen from Fig.\,\ref{fig4}, the SEA-obtained peak values are somewhat lower than the average peak values calculated based on Table \ref{tab_app1} (see Appendix \ref{app3}). This is of course due to the  SEA ``smoothing effect" (peaks do not occur at identical times). In Fig.\,\ref{fig5} the CR count, flow speed, magnetic field, and density SEA curves are overlaid (their peaks have been normalized to 1). As a result, their relative timings, duration, and changes are better visible. We note that for single events there might be some uncertainty in the timing of the zero epoch (onset time) due to manual detection, as described in Section \ref{data}. However, as can be seen in Fig.\,\ref{fig4}, these uncertainties are not systematic and the SEA flow speed increase starts at zero epoch. We note that the obtained SEA profiles visually reflects well the assumed structure of CIR according to the widely used sketch by \citet{belcher71}. According to this sketch. there are four consecutive regions of solar wind in a CIR: 1) the unperturbed slow solar wind, 2) the compressed and accelerated slow solar wind, 3) the compressed and decelerated fast solar wind, and 4) the unperturbed fast solar wind. The obtained SEA profiles and SEA-obtained peak values can be used as reference for modeling purposes.

We can see in Fig.\,\ref{fig5} that the duration of the plasma flow speed increase matches the duration of the recurrent FD well. The increase in flow speed shows a somewhat asymmetric profile compared to the CR count profile. This is associated to the differences observed in the declining phase of the time-profiles of both parameters, as shown in Fig.\,\ref{fig2}. Figure\,\ref{fig5} additionally shows that the magnetic field is enhanced during the main phase of the recurrent FD. There seems to be a small increase in CR count, which starts before the onset time (`the pre-increase''). Interestingly, the plasma density also starts to increase before the onset time. Perhaps less prominent, but visible, is the increase of $B$, which also starts before the onset time. Finally, there also seems to be a small decrease in the flow speed prior to the onset time. We note that all of the above might be related to the observed pre-increase of the CR count.

\section{Summary and conclusions}
\label{conclusion}

We studied a long-lived CIR originating from a single CH, recurring in 27 consecutive Carrington rotations (CR 2057--2083) in the time period from June 2007--May 2009. We analyzed the in situ measurements of this long-lived CIR, as well as the corresponding depression in the CR count observed by SOHO/EPHIN throughout the different rotations. 


We first analyzed the correlation between the plasma flow speed and mirrored CR count time-profiles and find them to be correlated, in agreement with previous studies. The correlation is different for different rotations and ranges from moderate to strong (Pearsons correlation coefficient between 0.57 and 0.95). In addition, we performed a statistical analysis and find that the CR count amplitude is correlated to the peak in the magnetic field and flow speed, as expected based on previous statistical studies. These results indicate that the combined convective-diffusion approach should be used in modeling of GCR modulation by CIRs.

In our study we use a ``clean'' sample based on a long-lived CIR, associated with a single CH. However, qualitatively, our results do not differ from those using samples related to different CHs. This indicates that the change of the physical properties of the recurring CIR from one rotation to another is qualitatively no different from the change of the physical properties of CIRs originating from different CHs.

Finally, in order to characterize a generic CIR and recurrent FD profile, we performed the superposed epoch analysis (SEA) using relative values of the key parameters. Thus, we obtained SEA profiles and SEA-obtained peak values that can be taken as a generic CIR example for modeling purposes.

\begin{acknowledgements}
B.V. and M.D. acknowledge support by the Croatian Science Foundation under the project IP-2020-02-9893 (ICOHOSS). BH. acknowledge the financial support via projects HE 3279/15-1 funded by the Deutsche Forschungsgemeinschaft (DFG). We acknowledge the SPDF COHOWeb and OMNIWeb databases as the source of data used, as well as the SOHO/EPHIN. SOHO/EPHIN is supported by the Ministry of Economics via DLR grant 50OG1702. The OMNI data were obtained from the GSFC/SPDF OMNIWeb interface at https://omniweb.gsfc.nasa.gov.
\end{acknowledgements}
%
\bibliographystyle{aa} 
\bibliography{REFs.bib}

\begin{thebibliography}{41}
\expandafter\ifx\csname natexlab\endcsname\relax\def\natexlab#1{#1}\fi

\bibitem[{{Badruddin}(1996)}]{badruddin96}
{Badruddin}. 1996, Astrophys.~\&~Space~Sci., 246, 171

\bibitem[{{Badruddin} \& {Kumar}(2016)}]{badruddin16}
{Badruddin} \& {Kumar}, A. 2016, Solar~Phys., 291, 559

\bibitem[{{Belcher} \& {Davis}(1971)}]{belcher71}
{Belcher}, J.~W. \& {Davis}, Leverett, J. 1971, J.~Geophys.~Res., 76, 3534

\bibitem[{{Burlaga} {et~al.}(1981){Burlaga}, {Sittler}, {Mariani}, \&
  {Schwenn}}]{burlaga81}
{Burlaga}, L., {Sittler}, E., {Mariani}, F., \& {Schwenn}, R. 1981,
  J.~Geophys.~Res., 86, 6673

\bibitem[{{Cane}(2000)}]{cane00}
{Cane}, H.~V. 2000, Space~Sci.~Rev., 93, 55

\bibitem[{{Delaboudini{\`e}re} {et~al.}(1995){Delaboudini{\`e}re}, {Artzner},
  {Brunaud}, {Gabriel}, {Hochedez}, {Millier}, {Song}, {Au}, {Dere}, {Howard},
  {Kreplin}, {Michels}, {Moses}, {Defise}, {Jamar}, {Rochus}, {Chauvineau},
  {Marioge}, {Catura}, {Lemen}, {Shing}, {Stern}, {Gurman}, {Neupert},
  {Maucherat}, {Clette}, {Cugnon}, \& {van Dessel}}]{delaboudiniere95}
{Delaboudini{\`e}re}, J.~P., {Artzner}, G.~E., {Brunaud}, J., {et~al.} 1995,
  Solar~Phys., 162, 291

\bibitem[{{Dresing} {et~al.}(2009){Dresing}, {G{\'o}mez-Herrero}, {Heber},
  {M{\"u}ller-Mellin}, {Wimmer-Schweingruber}, \& {Klassen}}]{dresing09}
{Dresing}, N., {G{\'o}mez-Herrero}, R., {Heber}, B., {et~al.} 2009,
  Solar~Phys., 256, 409

\bibitem[{{Dumbovi{\'c}} {et~al.}(2012){Dumbovi{\'c}}, {Vr{\v s}nak}, {{\v
  C}alogovi{\'c}}, \& {{\v Z}upan}}]{dumbovic12b}
{Dumbovi{\'c}}, M., {Vr{\v s}nak}, B., {{\v C}alogovi{\'c}}, J., \& {{\v
  Z}upan}, R. 2012, Astron.~Astrophys., 538, A28

\bibitem[{{Gil} \& {Mursula}(2018)}]{gil18}
{Gil}, A. \& {Mursula}, K. 2018, J.~Geophys.~Res., 123, 6148

\bibitem[{{Golub} {et~al.}(2007){Golub}, {Deluca}, {Austin}, {Bookbinder},
  {Caldwell}, {Cheimets}, {Cirtain}, {Cosmo}, {Reid}, {Sette}, {Weber},
  {Sakao}, {Kano}, {Shibasaki}, {Hara}, {Tsuneta}, {Kumagai}, {Tamura},
  {Shimojo}, {McCracken}, {Carpenter}, {Haight}, {Siler}, {Wright}, {Tucker},
  {Rutledge}, {Barbera}, {Peres}, \& {Varisco}}]{golub07}
{Golub}, L., {Deluca}, E., {Austin}, G., {et~al.} 2007, Solar~Phys., 243, 63

\bibitem[{{G{\'o}mez-Herrero} {et~al.}(2009){G{\'o}mez-Herrero}, {Klassen},
  {M{\"u}ller-Mellin}, {Heber}, {Wimmer-Schweingruber}, \&
  {B{\"o}ttcher}}]{gomez-herrero09}
{G{\'o}mez-Herrero}, R., {Klassen}, A., {M{\"u}ller-Mellin}, R., {et~al.} 2009,
  J.~Geophys.~Res., 114, A05101

\bibitem[{{G{\'o}mez-Herrero} {et~al.}(2011){G{\'o}mez-Herrero}, {Malandraki},
  {Dresing}, {Kilpua}, {Heber}, {Klassen}, {M{\"u}ller-Mellin}, \&
  {Wimmer-Schweingruber}}]{gomez-herrero11}
{G{\'o}mez-Herrero}, R., {Malandraki}, O., {Dresing}, N., {et~al.} 2011,
  J.~Atmos.~Sol.~Terr.~Phys., 73, 551

\bibitem[{{Gosling} {et~al.}(1978){Gosling}, {Asbridge}, {Bame}, \&
  {Feldman}}]{gosling78}
{Gosling}, J.~T., {Asbridge}, J.~R., {Bame}, S.~J., \& {Feldman}, W.~C. 1978,
  J.~Geophys.~Res., 83, 1401

\bibitem[{{Gosling} \& {Pizzo}(1999)}]{gosling99}
{Gosling}, J.~T. \& {Pizzo}, V.~J. 1999, Space~Sci.~Rev., 89, 21

\bibitem[{{Heinemann} {et~al.}(2020){Heinemann}, {Jer{\v{c}}i{\'c}}, {Temmer},
  {Hofmeister}, {Dumbovi{\'c}}, {Vennerstrom}, {Verbanac}, \&
  {Veronig}}]{heinemann20}
{Heinemann}, S.~G., {Jer{\v{c}}i{\'c}}, V., {Temmer}, M., {et~al.} 2020,
  Astron.~Astrophys., 638, A68

\bibitem[{{Heinemann} {et~al.}(2019){Heinemann}, {Temmer}, {Heinemann},
  {Dissauer}, {Samara}, {Jer{\v{c}}i{\'c}}, {Hofmeister}, \&
  {Veronig}}]{heinemann19b}
{Heinemann}, S.~G., {Temmer}, M., {Heinemann}, N., {et~al.} 2019, Solar~Phys.,
  294, 144

\bibitem[{{Hofmeister} {et~al.}(2018){Hofmeister}, {Veronig}, {Temmer},
  {Vennerstrom}, {Heber}, \& {Vr{\v s}nak}}]{hofmeister18}
{Hofmeister}, S.~J., {Veronig}, A., {Temmer}, M., {et~al.} 2018,
  J.~Geophys.~Res., 123, 1738

\bibitem[{{Jian} {et~al.}(2006){Jian}, {Russell}, {Luhmann}, \&
  {Skoug}}]{jian06b}
{Jian}, L., {Russell}, C.~T., {Luhmann}, J.~G., \& {Skoug}, R.~M. 2006,
  Solar~Phys., 239, 337

\bibitem[{{Kilpua} {et~al.}(2017){Kilpua}, {Koskinen}, \&
  {Pulkkinen}}]{kilpua17}
{Kilpua}, E., {Koskinen}, H.~E.~J., \& {Pulkkinen}, T.~I. 2017, Living Reviews
  in Solar Physics, 14, 5

\bibitem[{{King} \& {Papitashvili}(2005)}]{king05}
{King}, J.~H. \& {Papitashvili}, N.~E. 2005, J.~Geophys.~Res., 110, A02104

\bibitem[{{Kosugi} {et~al.}(2007){Kosugi}, {Matsuzaki}, {Sakao}, {Shimizu},
  {Sone}, {Tachikawa}, {Hashimoto}, {Minesugi}, {Ohnishi}, {Yamada}, {Tsuneta},
  {Hara}, {Ichimoto}, {Suematsu}, {Shimojo}, {Watanabe}, {Shimada}, {Davis},
  {Hill}, {Owens}, {Title}, {Culhane}, {Harra}, {Doschek}, \&
  {Golub}}]{kosugi07}
{Kosugi}, T., {Matsuzaki}, K., {Sakao}, T., {et~al.} 2007, Solar~Phys., 243, 3

\bibitem[{{K{\"u}hl} {et~al.}(2015){K{\"u}hl}, {Banjac}, {Heber}, {Labrenz},
  {M{\"u}ller-Mellin}, \& {Terasa}}]{kuhl15}
{K{\"u}hl}, P., {Banjac}, S., {Heber}, B., {et~al.} 2015,
  Cent.~European~Astrophys.~Bull., 39, 119

\bibitem[{{K{\"u}hl} {et~al.}(2013){K{\"u}hl}, {Dresing}, {Dunzlaff},
  {Fichtner}, {Gieseler}, {G{\'o}mez-Herrero}, {Heber}, {Klassen}, {Kleimann},
  {Kopp}, {Potgieter}, {Scherer}, \& {Strauss}}]{kuhl13}
{K{\"u}hl}, P., {Dresing}, N., {Dunzlaff}, P., {et~al.} 2013,
  Cent.~European~Astrophys.~Bull., 37, 643

\bibitem[{{Kumar} \& {Badruddin}(2014)}]{kumar14}
{Kumar}, A. \& {Badruddin}. 2014, Solar~Phys., 289, 2177

\bibitem[{{Lopez}(1987)}]{lopez87}
{Lopez}, R.~E. 1987, J.~Geophys.~Res., 92, 11189

\bibitem[{{Melkumyan} {et~al.}(2019){Melkumyan}, {Belov}, {Abunina}, {Abunin},
  {Eroshenko}, {Yanke}, \& {Oleneva}}]{melkumyan19}
{Melkumyan}, A.~A., {Belov}, A.~V., {Abunina}, M.~A., {et~al.} 2019,
  Adv.~Space~Res., 63, 1100

\bibitem[{{M{\"u}ller} {et~al.}(2017){M{\"u}ller}, {Nicula}, {Felix},
  {Verstringe}, {Bourgoignie}, {Csillaghy}, {Berghmans}, {Jiggens},
  {Garc{\'\i}a-Ortiz}, {Ireland}, {Zahniy}, \& {Fleck}}]{muller17}
{M{\"u}ller}, D., {Nicula}, B., {Felix}, S., {et~al.} 2017, Astron.~Astrophys.,
  606, A10

\bibitem[{{M{\"u}ller-Mellin} {et~al.}(1995){M{\"u}ller-Mellin}, {Kunow},
  {Flei{\ss}ner}, {Pehlke}, {Rode}, {R{\"o}schmann}, {Scharmberg}, {Sierks},
  {Rusznyak}, {McKenna-Lawlor}, {Elendt}, {Sequeiros}, {Meziat}, {Sanchez},
  {Medina}, {Del Peral}, {Witte}, {Marsden}, \& {Henrion}}]{muller-mellin95}
{M{\"u}ller-Mellin}, R., {Kunow}, H., {Flei{\ss}ner}, V., {et~al.} 1995,
  Solar~Phys., 162, 483

\bibitem[{{Neugebauer} {et~al.}(2002){Neugebauer}, {Liewer}, {Smith}, {Skoug},
  \& {Zurbuchen}}]{neugebauer02}
{Neugebauer}, M., {Liewer}, P.~C., {Smith}, E.~J., {Skoug}, R.~M., \&
  {Zurbuchen}, T.~H. 2002, J.~Geophys.~Res., 107, 1488

\bibitem[{{Nolte} {et~al.}(1976){Nolte}, {Krieger}, {Timothy}, {Gold},
  {Roelof}, {Vaiana}, {Lazarus}, {Sullivan}, \& {McIntosh}}]{nolte76}
{Nolte}, J.~T., {Krieger}, A.~S., {Timothy}, A.~F., {et~al.} 1976, Solar~Phys.,
  46, 303

\bibitem[{{Parker}(1965)}]{parker65}
{Parker}, E.~N. 1965, Planet.~Space.~Sci., 13, 9

\bibitem[{{Richardson}(2004)}]{richardson04}
{Richardson}, I.~G. 2004, Space~Sci.~Rev., 111, 267

\bibitem[{{Richardson}(2018)}]{richardson18}
{Richardson}, I.~G. 2018, Living Reviews in Solar Physics, 15, 1

\bibitem[{{Richardson} \& {Cane}(1995)}]{richardson95}
{Richardson}, I.~G. \& {Cane}, H.~V. 1995, J.~Geophys.~Res., 100, 23397

\bibitem[{{Richardson} \& {Cane}(2010)}]{richardson10}
{Richardson}, I.~G. \& {Cane}, H.~V. 2010, Solar~Phys., 264, 189

\bibitem[{{Richardson} {et~al.}(1996){Richardson}, {Wibberenz}, \&
  {Cane}}]{richardson96}
{Richardson}, I.~G., {Wibberenz}, G., \& {Cane}, H.~V. 1996, J.~Geophys.~Res.,
  101, 13483

\bibitem[{{Tokumaru} {et~al.}(2017){Tokumaru}, {Satonaka}, {Fujiki}, {Hayashi},
  \& {Hakamada}}]{tokumaru17}
{Tokumaru}, M., {Satonaka}, D., {Fujiki}, K., {Hayashi}, K., \& {Hakamada}, K.
  2017, Solar~Phys., 292, 41

\bibitem[{{{\v C}alogovi{\'c}} {et~al.}(2009){{\v C}alogovi{\'c}}, {Vr{\v
  s}nak}, {Temmer}, \& {Veronig}}]{calogovic09}
{{\v C}alogovi{\'c}}, J., {Vr{\v s}nak}, B., {Temmer}, M., \& {Veronig}, A.~M.
  2009, in IAU~Symp., Vol. 257, IAU~Symp., ed. {N.~Gopalswamy \& D.~F.~Webb},
  425--427

\bibitem[{{Vr{\v s}nak} {et~al.}(2007){Vr{\v s}nak}, {Temmer}, \&
  {Veronig}}]{vrsnak07a}
{Vr{\v s}nak}, B., {Temmer}, M., \& {Veronig}, A.~M. 2007, Solar~Phys., 240,
  315

\bibitem[{{Vr{\v{s}}nak} {et~al.}(2022){Vr{\v{s}}nak}, {Dumbovi{\'c}}, {Heber},
  \& {Kirin}}]{vrsnak22}
{Vr{\v{s}}nak}, B., {Dumbovi{\'c}}, M., {Heber}, B., \& {Kirin}, A. 2022,
  Astron.~Astrophys., In Press

\bibitem[{{Zurbuchen} \& {Richardson}(2006)}]{zurbuchen06}
{Zurbuchen}, T.~H. \& {Richardson}, I.~G. 2006, Space~Sci.~Rev., 123, 31

\end{thebibliography}
%
\begin{appendix} 
\section{Key parameter measurements}
\label{app1}
Overview of normalization parameters and peak values for all rotations of the recurring CIR are given in Table \ref{tab_app1}.
\begin{table}[h]
\caption{CIR normalization parameters and peak values throughout different rotations.}
\label{tab_app1}
\centering
\begin{tabular}{ccccccccccccc} 
\hline    
                &       $Brms$  &       $Brms$  &       $B$     &       $B$     &       $v$                                     &       $v$     &       $n$                                     &       $n$     &       $T$                             &       $T$     &                               &       $CR\,\,count$\\
rotation        &       norm            &       peak            &       norm    &       peak    &       norm                                    &       peak    &       norm                                    &       peak    &       norm                            &       peak    &       $CR\,\,count$   &       peak\\
number  &       [nT]            &        [\%]           &        [nT]   &        [\%]    &        [$\mathrm{km~s}^{-1}$] &        [\%]   &       [$\mathrm{cm}^{-3}$]    &        [\%]    &        [$\cdot10^{5}$\,K]     &        [\%]   &       norm                    &       [\%]\\
\hline
rot1    &       0,12    &       1357    &       2,7     &       244     &       327     &       38      &       4,5     &       371     &       0,31    &       396     &       22729   &       1,3     \\
rot2    &       0,25    &       900     &       3,0     &       249     &       346     &       82      &       3,9     &       360     &       0,43    &       738     &       22940   &       1,1     \\
rot3    &       0,35    &       643     &       2,5     &       532     &       367     &       77      &       4,1     &       463     &       0,45    &       594     &       22968   &       2,6     \\
rot4    &       0,40    &       425     &       2,4     &       720     &       311     &       117     &       8,0     &       484     &       0,16    &       2744    &       23594   &       5,5     \\
rot5    &       0,35    &       557     &       3,1     &       239     &       337     &       102     &       8,0     &       399     &       0,34    &       1196    &       23470   &       3,6     \\
rot6    &       0,15    &       1567    &       3,3     &       288     &       321     &       109     &       5,2     &       165     &       0,25    &       1660    &       23598   &       3,9     \\
rot7    &       0,13    &       1925    &       4,0     &       214     &       369     &       81      &       3,3     &       497     &       0,53    &       564     &       23376   &       2,7     \\
rot8    &       0,13    &       2000    &       3,3     &       365     &       302     &       111     &       6,9     &       214     &       0,21    &       1942    &       24325   &       6,3     \\
rot9    &       0,08    &       4340    &       2,0     &       677     &       311     &       129     &       7,9     &       393     &       0,10    &       4428    &       24522   &       10,3    \\
rot10   &       0,12    &       1700    &       2,9     &       289     &       339     &       86      &       3,1     &       1049    &       0,46    &       582     &       24017   &       3,7     \\
rot11   &       0,23    &       1014    &       3,4     &       225     &       342     &       129     &       3,9     &       417     &       0,37    &       1217    &       23573   &       2,6     \\
rot12   &       0,43    &       454     &       2,9     &       224     &       361     &       90      &       3,0     &       827     &       0,65    &       490     &       23379   &       2,4     \\
rot13   &       0,20    &       700     &       1,7     &       737     &       363     &       81      &       4,0     &       382     &       0,31    &       2265    &       23713   &       3,7     \\
rot14   &       0,15    &       767     &       3,0     &       178     &       329     &       83      &       4,2     &       493     &       0,28    &       896     &       23722   &       4,3     \\
rot15   &       0,32    &       847     &       2,1     &       616     &       312     &       144     &       7,9     &       382     &       0,14    &       4860    &       23836   &       5,0     \\
rot16   &       0,25    &       1100    &       3,7     &       305     &       347     &       104     &       7,5     &       183     &       0,35    &       939     &       23877   &       4,7     \\
rot17   &       0,20    &       2100    &       3,1     &       523     &       334     &       97      &       3,8     &       689     &       0,44    &       909     &       23994   &       3,0     \\
rot18   &       0,27    &       687     &       3,5     &       314     &       298     &       112     &       6,1     &       308     &       0,38    &       692     &       24707   &       4,5     \\
rot19   &       0,18    &       664     &       4,5     &       87      &       342     &       109     &       1,8     &       394     &       0,40    &       770     &       24669   &       3,2     \\
rot20   &       0,17    &       1100    &       5,0     &       171     &       335     &       112     &       3,0     &       335     &       0,81    &       364     &       24850   &       3,8     \\
rot21   &       0,15    &       1900    &       2,2     &       903     &       314     &       110     &       7,3     &       893     &       0,12    &       3913    &       25398   &       3,8     \\
rot22   &       0,12    &       929     &       4,9     &       137     &       295     &       89      &       3,4     &       640     &       0,49    &       370     &       25108   &       1,5     \\
rot23   &       0,13    &       1550    &       4,0     &       204     &       318     &       51      &       2,8     &       293     &       0,37    &       683     &       25754   &       2,6     \\
rot24   &       0,18    &       1264    &       2,7     &       496     &       300     &       96      &       3,5     &       2014    &       0,43    &       884     &       26378   &       3,7     \\
rot25   &       0,33    &       890     &       2,6     &       593     &       319     &       77      &       7,3     &       201     &       0,39    &       924     &       26606   &       2,7     \\
rot26   &       0,32    &       532     &       2,9     &       190     &       311     &       80      &       4,0     &       432     &       0,32    &       629     &       26867   &       2,0     \\
rot27   &       0,22    &       685     &       1,6     &       345     &       334     &       56      &       3,2     &       407     &       0,28    &       504     &       27154   &       2,3     \\
\hline
\end{tabular}
\end{table}
\clearpage
\section{ICME signatures in rotations 1, 18, and 25}
\label{app2}

\begin{figure}[h]
\centering
\includegraphics[width=0.48\textwidth]{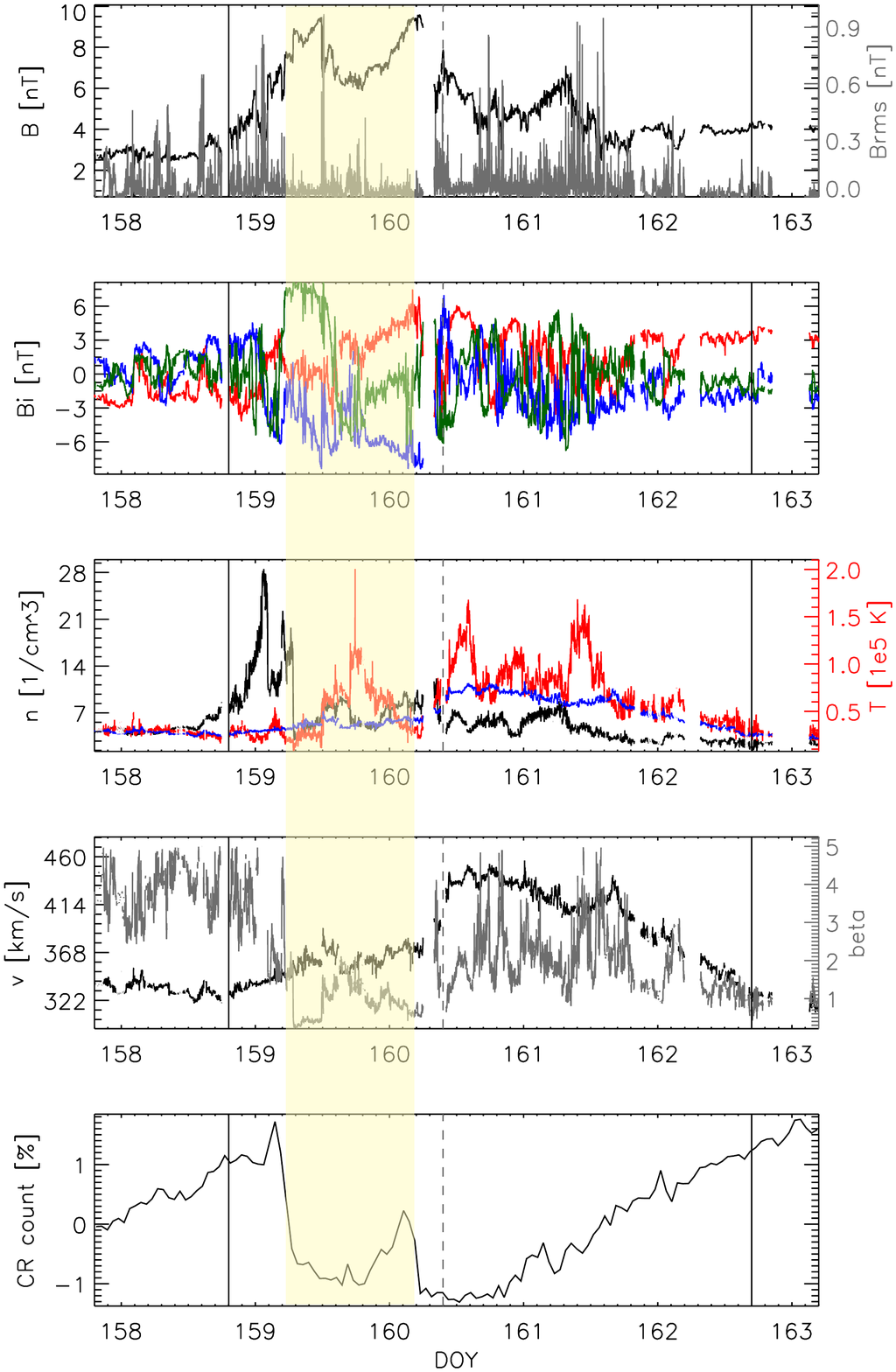}
\caption{In situ measurements given in day-of-year (DOY) time series of 2007 corresponding to rotation 1. The ICME signatures are highlighted yellow (for other details see main text).}
\label{fig_app2a}
\end{figure}

Figs.\,\ref{fig_app2a}--\,\ref{fig_app2c} show 1 minute plasma and magnetic field data in the Geocentric solar ecliptic (GSE) system provided by the \href{https://omniweb.gsfc.nasa.gov/html/ow_data.html#norm_pla}{OMNIWeb} database \citep{king05} for rotations 1, 18, and 25. Panels top to bottom show: 1) magnetic field strength (black) and its fluctuations, $dBrms$; 2) magnetic field GSE components x, y, and z colored red, blue and green, respectively; 3) plasma density (black), temperature (red) and expected temperature (blue); 4) plasma flow speed (black) and beta (gray); 5) the SOHO/EPHIN F-detector particle count, $CR\,\,count$. The expected temperature was calculated according to \citet{lopez87} and \citet{richardson95}. The start and end time of the analysed time interval are defined as for the analysis of 1-hour in situ data for the corresponding rotation. Solid lines mark the onset and trailing edge time of the CIR (if present), whereas the dashed gray line marks the stream interface time (as given in Table\ref{tab1}).

It can be seen in Fig.\,\ref{fig_app2a} that from DOY 159.3 until DOY 160.3 (highlighted yellow), the magnetic field components show smoother profile than in the rest of the observed time-interval. In the highlighted region we observe a rotation of $B_z$, and at the start of the highlighted region temperature and plasma beta parameter are decreased. These are indications of an ICME, that is, more specifically of a magnetic cloud \citep[MC][]{burlaga81,zurbuchen06,kilpua17}. However, the plasma flow speed does not show the monotonically declining profile characteristic for MCs, which is related to their expansion. This is most probably due to inhibited expansion, as the ICME is ``pushed" from behind by the HSS. The HSS after the trailing edge of the ICME is identified as increase of plasma flow speed and temperature. In the trailing part of the highlighted region we observe increased temperature and plasma beta, as well as a slightly increasing plasma flow speed profile, which is most likely due to the acceleration from the faster HSS in the back. During the whole observed time interval, two depressions can be observed in the CR count, where the first depression roughly corresponds to the highlighted region and is thus most likely an ICME-related Forbush decrease. The second depression starts slightly before the stream interface and lasts throughout the HSS, similarly as recurrent Forbush decreases observed in other rotations where ICME signatures are not observed.
\begin{figure}[h]
\centering
\includegraphics[width=0.48\textwidth]{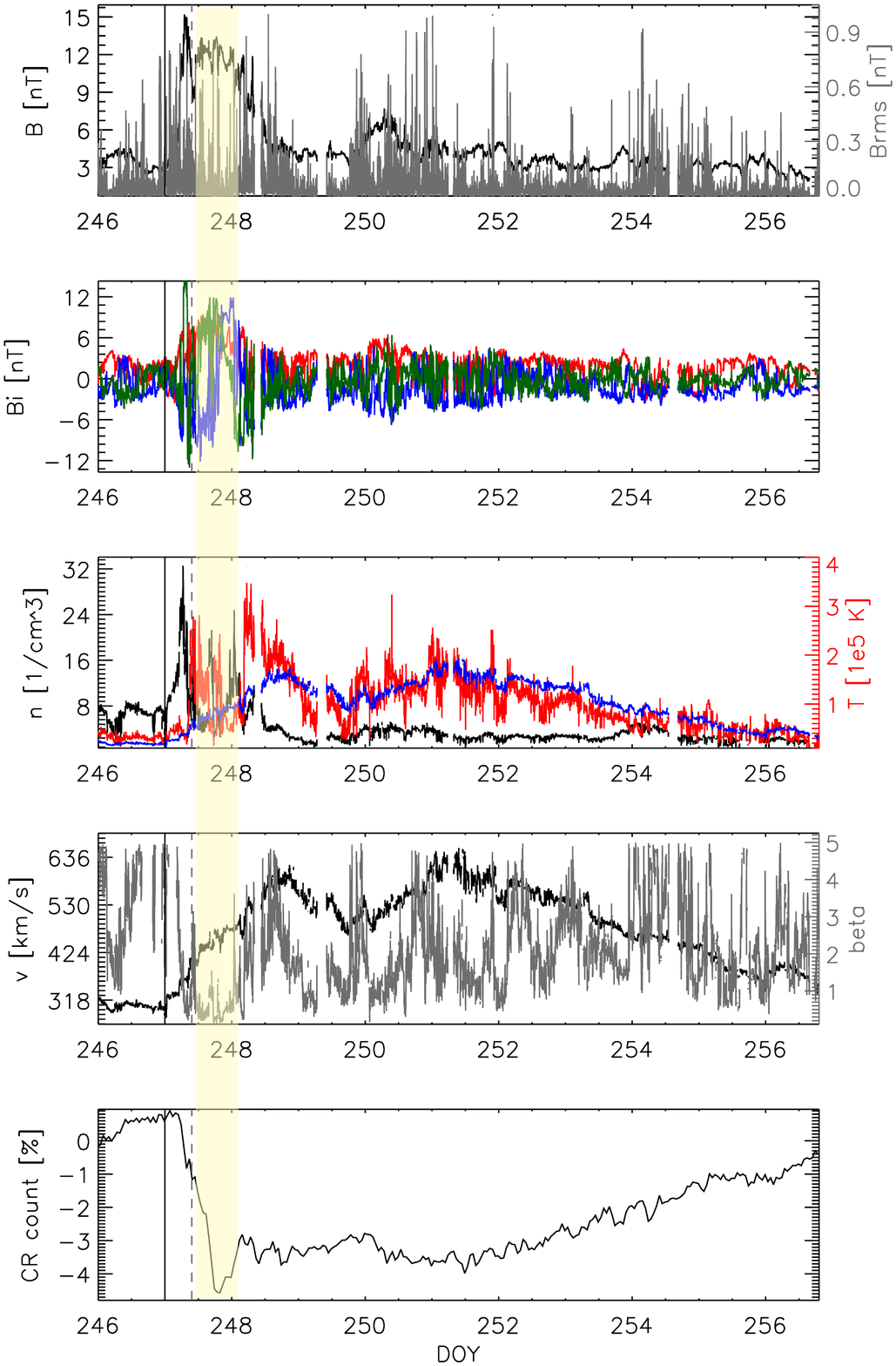}
\caption{In situ measurements given in day-of-year (DOY) time series of 2008 corresponding to rotation 18. The ICME signatures are highlighted yellow (for other details see main text).}
\label{fig_app2b}
\end{figure}

In Fig.\,\ref{fig_app2b}, we observe rotation in the $B_y$ from DOY 247.5 until DOY 248.1 (highlighted yellow), accompanied by a decrease of plasma beta and an additional decrease in CR count, superimposed to the decrease that has already started. However, the plasma density and temperature are not decreased. These are indications of the ejecta type of ICMEs \citep[i.e., without clear MC signatures][]{richardson10,kilpua17}. The plasma flow speed shows a monotonically increasing plasma flow speed profile, indicating acceleration of plasma flow throughout the structure. The most likely scenario explaining these signatures is that the HSS is pushing and accelerating the ICME from the back. We note that we observe a double peak in the plasma flow speed profile, corresponding to the double peak in the plasma temperature. This is probably related to the double-HSS structure, as the corresponding CH shows an inhomogeneous, ``patchy'' structure (determined visually using the \href{https://www.solarmonitor.org}{Solar Monitor} service for the rotation 18 CH date as given in Table\ref{tab1}, not shown here). We note that ICME-related depression in the CR count seems superimposed to the larger-scale CIR-related depression, unlike the event shown in Fig.\,\ref{fig_app2a}, where two clear and separate depressions can be identified.

\begin{figure}[h]
\centering
\includegraphics[width=0.48\textwidth]{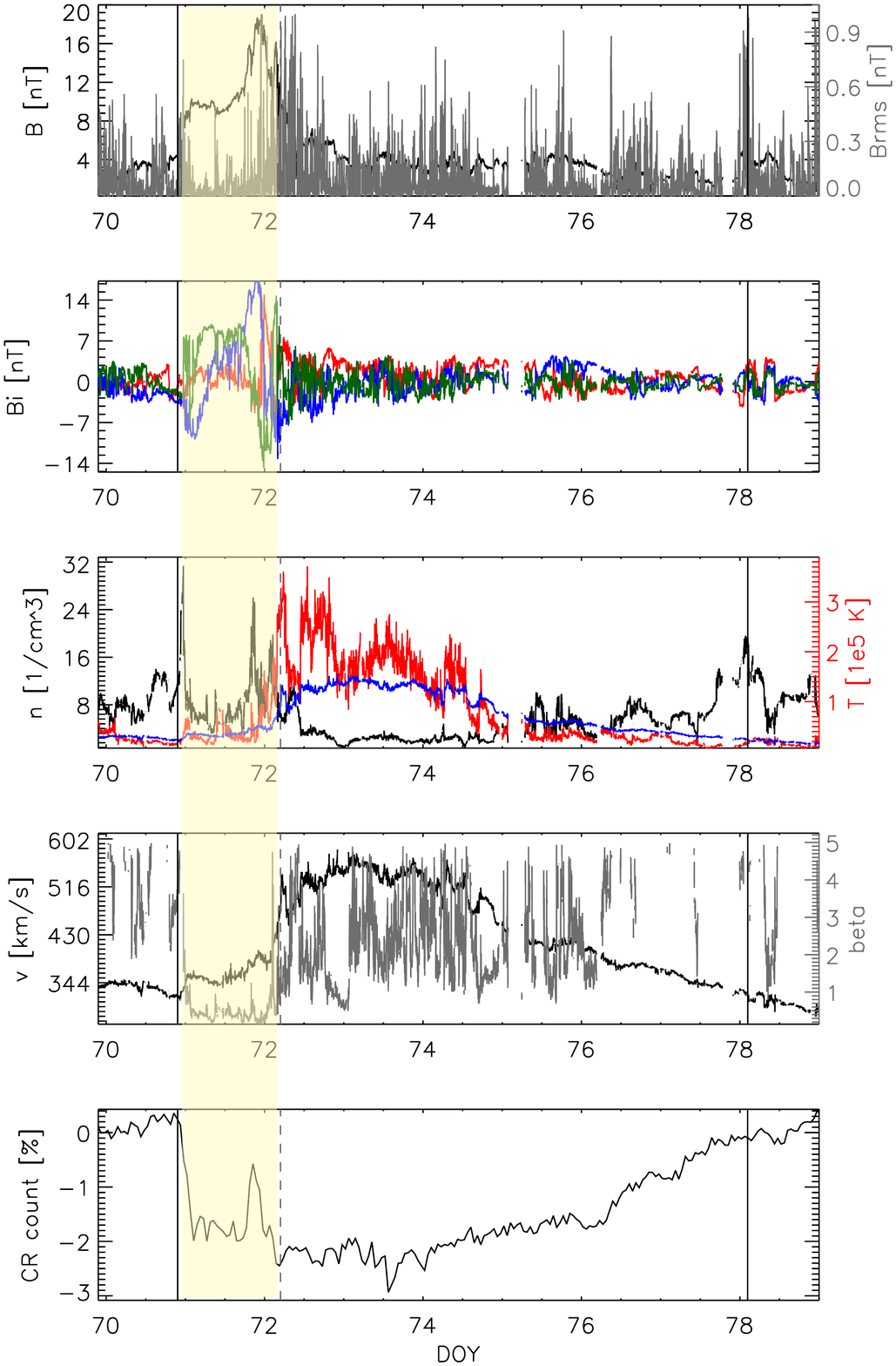}
\caption{In situ measurements given in day-of-year (DOY) time series of 2009 corresponding to rotation 25. The ICME signatures are highlighted yellow (for other details see main text).}
\label{fig_app2c}
\end{figure}

In Fig.\,\ref{fig_app2c}, we observe smooth rotation in the $B_y$ from DOY 71 until roughly DOY 72, accompanied by a decrease of plasma beta (highlighted yellow) and occasional decreases of temperature below expected temperature. As in previous two events, these are indications of an ICME \citep{zurbuchen06,kilpua17}. Density is increased throughout this time period and the plasma flow speed profile does not show a monotonically declining (expanding) profile. On the contrary, the flow speed profile is slightly increasing towards the ICME trailing edge. Therefore, again in this case the ICME seems to be pushed and accelerated from the back by the HSS. As in the case of event shown in Fig.\,\ref{fig_app2a}, two depressions can be observed in the CR count. The first depression is likely related to the ICME and the second to the HSS.

It is interesting to note that in all three observed cases of ICME-CIR interaction, the ICME expansion is inhibited and ICME seems to be pushed (and accelerated) from the HSS behind. In all three cases, the ICME is slower than the HSS. However, the location of the stream interface seems to be different for event 2 shown in  Fig.\,\ref{fig_app2b} compared to other two events. More specifically, in the former case (event 2), it is located in front of the ICME whereas in the other two events it is located after the ICME. We also note that in event 2, we see a fully accelerating plasma flow speed profile throughout the ICME, whereas in the other two events, only the trailing part of the ICME plasma seems to be accelerated. Therefore, it is possible that these three examples show different stages of the evolution of the ICME-CIR interaction. In events 1 and 3, the interaction could be at an early stage, during which the HSS stopped the ICME expansion and started to accelerate its plasma flow from behind. In event 2, the interaction could already be at an advanced stage, where the plasma flow inside the ICME is already accelerated up to a point that the ICME is accelerating the plasma of the slow solar wind in front, thus creating (i.e., ``shifting'') the stream interface from its back to the front. Additional support for this different-stage-of-evolution interpretation is the fact that events 1 and 3 have a more typical ICME duration \citep[i.e., $\approx 24$hours][]{richardson10} and clearer ICME signatures compared to event 2, which has much shorter duration (14 hours).

\clearpage
\section{CIRs in rotations 9 and 24 -- the outliers}
\label{app3}

\begin{figure}[h]
\centering
\includegraphics[width=0.48\textwidth]{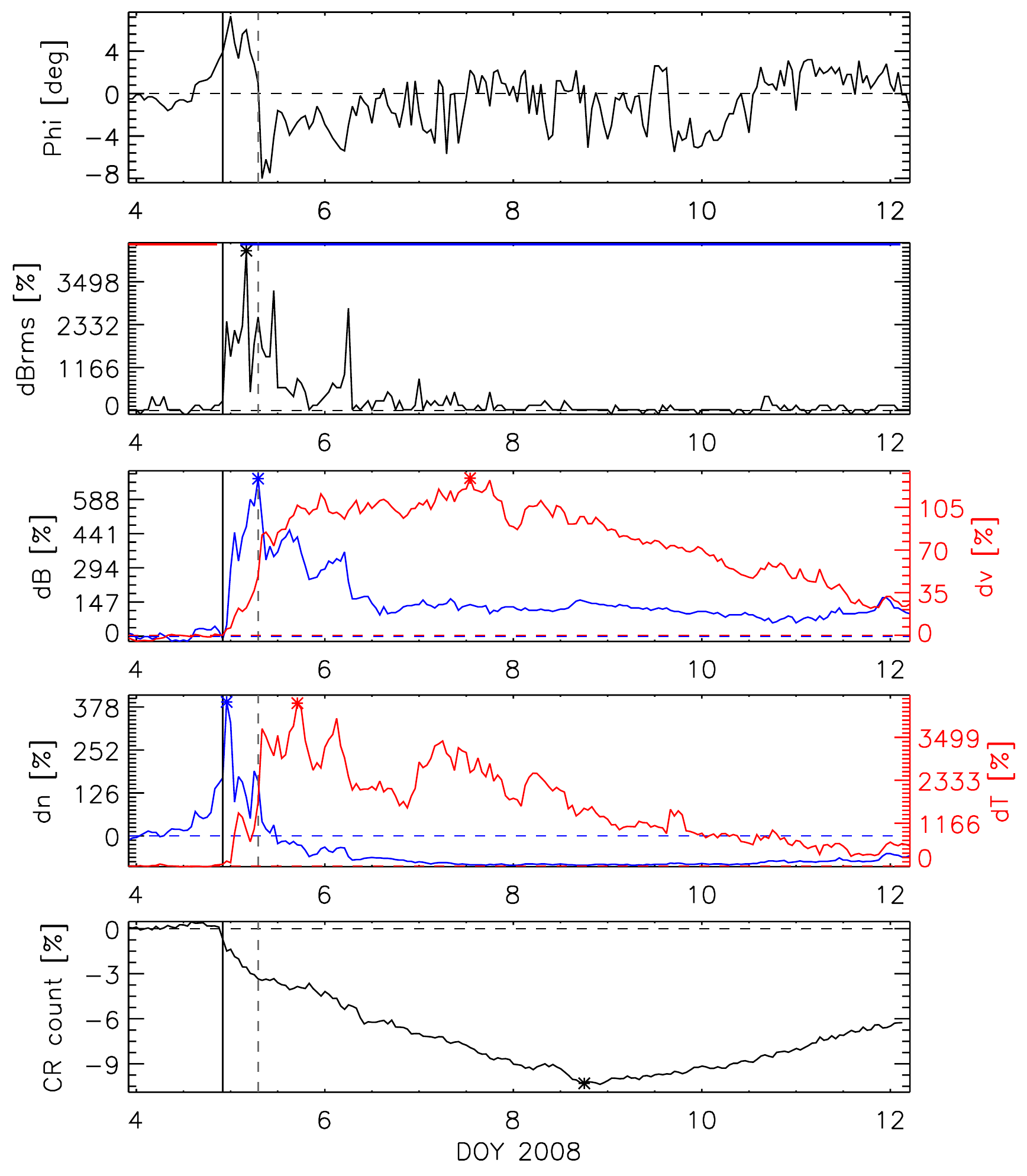}
\caption{In situ measurements for CIR in carrington rotation 2065 (rot9). For details see main text.}
\label{fig_app3a}
\end{figure}

\begin{figure}[h]
\centering
\includegraphics[width=0.48\textwidth]{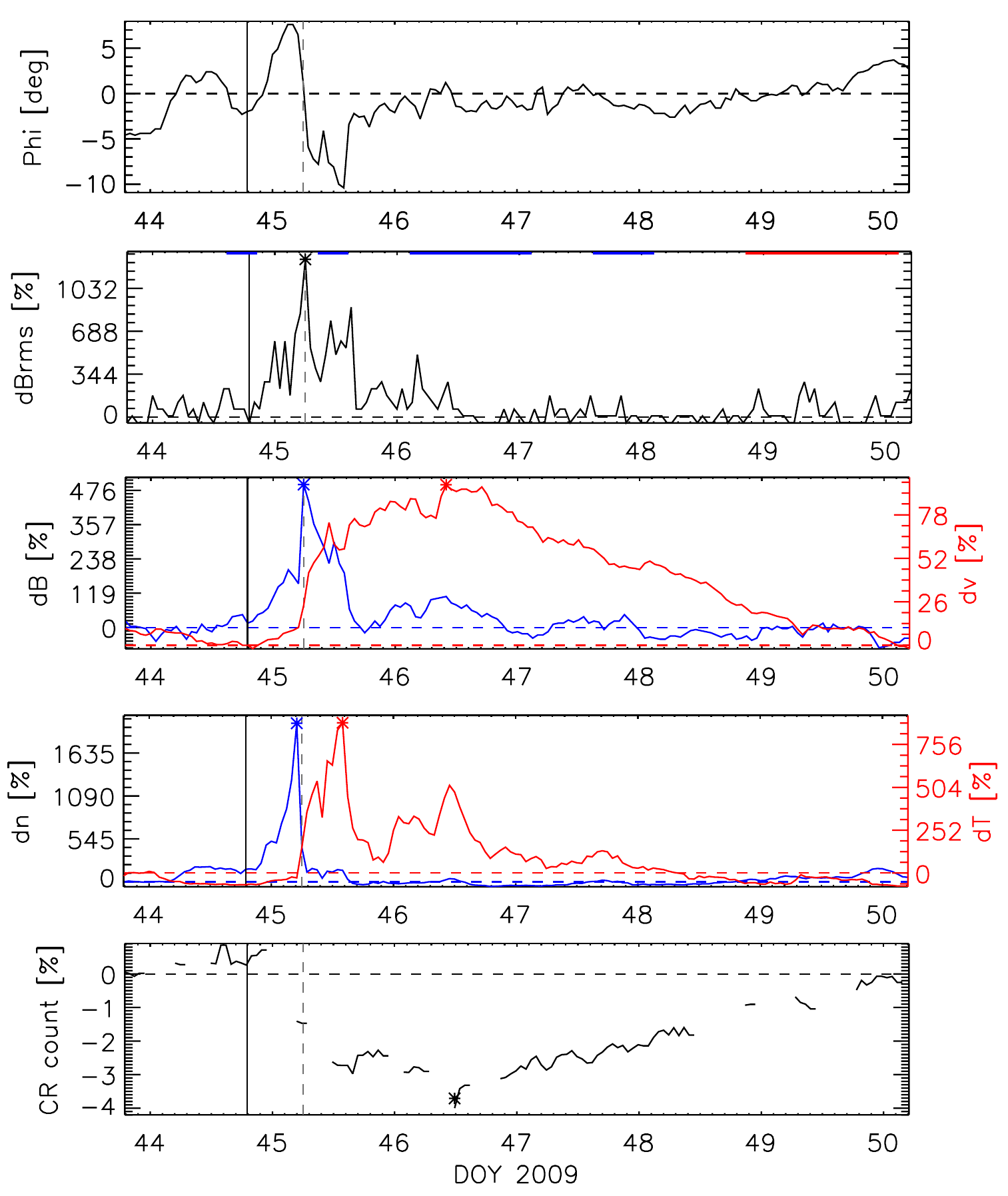}
\caption{In situ measurements for CIR in carrington rotation 2080 (rot24). For details see main text.}
\label{fig_app3b}
\end{figure}

\begin{table}[h]
\fontsize{6}{12}\selectfont
\caption{Normalization parameters and peak values: outliers versus average values}
\label{tab_app2}
\centering
\begin{tabular}{ccccccc} 
\hline    
in                                              &       average                 &       rot9            &       rot24           &       average                 &       rot9            &       rot24   \\
situ                                            &       pre-event               &       pre-event       &       pre-event       &       peak                    &       peak            &       peak    \\
parameter                                       &       values          &       value   &       value   &       values          &       value   &       value\\
\hline
$Brms$ [nT]                             &       $0.22\pm0.10$   &       0.08            &       0.18            &       $2.40\pm0.73$   &       3.70            &       2.50    \\
$B$ [nT]                                        &       $3.1\pm0.9$     &       2.0             &       2.7             &       $13.2\pm3.8$    &       15.8            &       16.1    \\
$v$ [$\mathrm{km~s}^{-1}$]      &       $329\pm21$      &       311             &       300             &       $639\pm79$      &       712             &       588     \\
$n$ [$\mathrm{cm}^{-3}$]                &       $4.9\pm2.0$     &       7.9             &       3.5             &       $27.7\pm16.0$   &       38.9            &       74.0    \\
$T$ [$\cdot10^{5}$\,K]          &       $0.36\pm0.16$   &       0.1             &       0.43            &       $3.83\pm1.30$   &       4.44            &       4.23    \\
$CR\,count$ [\%]                        &       --                      &       --              &       --              &       $3.6\pm1.8$     &       10.3            &       3.7     \\                                                                                                      
\hline
\end{tabular}
\end{table}

Figures\,\ref{fig_app3a}--\,\ref{fig_app3b} show in situ measurements for CIRs in carrington rotation 2065 (rot9) and 2080 (rot24). The panels top to bottom show: 1) the azimuth flow angle calculated in the RTN system, $Phi$; 2) the magnetic field fluctuations, $dBrms$, and the in situ magnetic polarity (red and blue overlying lines for positive and negative polarity, respectively, with details in Section \ref{data};  3) the total magnetic field strength, $B$, and plasma flow speed, $v$; 4) plasma density, $n$, and plasma temperature, $T$; and 5) the SOHO/EPHIN F-detector particle count, $CR\,\,count$. Vertical solid lines mark the onset and the trail edge times, whereas the vertical dashed line marks the stream interface time. All parameters, except $Phi$ are given in relative values. Horizontal dashed lines mark reference levels, which were obtained by normalising. Asterisks mark the measured peak values. The comparison of normalization parameters and peak values of events in rotation 9 and 24 with average values for all 27 rotations is given in Table \ref{tab_app2}.

The outlier marked red in Fig.\,\ref{fig3} (rot9) is shown in Fig.\,\ref{fig_app3a}. As can be seen in Table \ref{tab_app2}, the pre-event values for $Brms$, $B$, $v$ and $T$ are quite low, whereas the pre-event value for $n$ is quite high. On the other hand, the peak values corresponding to this event are relatively high. We note that the maximum peak values in the sample do not correspond to this event for any of the parameters, however, the event does have a minimum normalization values in the whole sample of 27 rotations for $Brms$ and $T$. We also note a very narrow and pronounced peak of $Brms$ almost two times higher than the rest of the curve in the period of increased $Brms$. The duration of the event (from onset time to end time, see Table \ref{tab1}) is 7.3 days, which is only slightly longer than the average ($6.8\pm1.5$days). The duration of the compression region upstream of the stream interface (SI time-onset time, see Table \ref{tab1}) is only slightly shorter than the average ($0.6\pm0.4$days). We note that, compared to other events, the flow speed reaches its peak relatively late after the stream interface and it never returns to the pre-event level. In addition, the magnetic field remains high throughout the event and does not return to the pre-event level. However, we also observed such behavior in some other rotations as well (see Fig.\,\ref{fig2}); therefore, according to its profile, this particular rotation does not seem to be an outlier. It is possible that rot9 only appears as an outlier in the scatter-plots due to low normalization factor and extreme narrow peak of $Brms$.

The outlier marked blue in Fig.\,\ref{fig3} (rot24) is shown in Fig.\,\ref{fig_app3b}. As can be seen in Table \ref{tab_app2}, the pre-event and peak values of all parameters for this event are quite normal, except for density. The peak value of density is the largest measured peak value in all of rotations (see Table \ref{tab_app1}). Therefore, rot24 only appears as an outlier in the density scatter plot, but it does not seem to posses any other extraordinary features.

\end{appendix}
\end{document}